\documentclass[12pt,draftclsnofoot, onecolumn,draftcls]{IEEEtran}
\usepackage{graphicx,cite,epsfig,amssymb,amsmath,subfigure,url,stfloats,latexsym}
\usepackage{multirow,array}
\usepackage{graphicx,cite,amsmath,amssymb,hhline,booktabs,stfloats,bm}
\usepackage{verbatim}
\usepackage[utf8]{inputenc}
\usepackage{subfigure}
\usepackage{algorithm,algorithmic}
\usepackage{color}
\usepackage{caption2}

\begin{document}

\title{\mbox{}\vspace{1.5cm}\\
\textsc{
Dynamic Load-Balancing Vertical Control for Large-Scale Software-Defined Internet of Things} \vspace{1.5cm}
}

\author{Lianming~Zhang,~Xiaoxun~Zhong,~Yehua Wei, and~Kun~Yang,~\IEEEmembership{Senior~Member,~IEEE}
\thanks{L. Zhang (zlm@hunnu.edu.cn), X. Zhong and Y. Wei (yehuahn@hunnu.edu.cn) are with Key Laboratory of Internet of Things Technology and Application, College of Physics and Information Science, Hunan Normal University, Changsha, China. K. Yang (kunyang@essex.ac.uk) is with the School of Computer Sciences and Electrical Engineering, University of Essex, CO4 3SQ, Colchester, UK.}
}

\date{\today}
\renewcommand{\baselinestretch}{1.2}
\thispagestyle{empty} \maketitle \thispagestyle{empty}

\begin{abstract}
As the global Internet of things increasingly is popular with consumers and business environment, network flow management has become an important topic to optimize the performance on Internet of Things. The rigid existing Internet of things (IoT) architecture blocks current traffic management technology to provide a real differentiated service for large-scale IoT. Software-defined Internet of Things (SD-IoT) is a new computing paradigm that separates control plane and data plane, and enables centralized logic control. In this paper, we first present a general framework for SD-IoT, which consists of two main components: SD-IoT controllers and SD-IoT switches. The controllers of SD-IoT uses resource pooling technology, and the pool is responsible for the centralized control of the entire network. The switches of SD-IoT integrate with the gateway functions, which is responsible for data access and forwarding. The SD-IoT controller pool is designed as a vertical control architecture, which includes the main control layer and the base control layer. The controller (main controller) of the main control layer interacts upward with the application layer, interacts with the base control layer downwards, and the controller (base controller) of the basic control layer interacts with the data forwarding layer. We propose a dynamic balancing algorithm of the main controller based on election mechanism and a dynamic load balancing algorithm of the basic controller based on the balanced delay, respectively. The experimental results show that the dynamic balancing algorithm based on the election mechanism can ensure the consistency of the messages between the main controllers, and the dynamic load balancing algorithm based on the balanced delay can balance between these different workloads in the basic controllers.
\end{abstract}


\IEEEpeerreviewmaketitle

\vspace{0.3in}
\section{Introduction}
IoT is a kind of the Internet which is connected to the things. IoT contains a large number of interconnected devices, including household appliance, public facilities, wearable equipments, residential office buildings, industrial processes, medical equipments, law enforcement equipments, military installations, unmanned aerial vehicles, interconnected cars and other network applications that may be almost impossible to imagine at present \cite{b1}. IoT provides a huge market opportunity for equipment manufacturers, Internet service suppliers and application development engineers. At present, the global application of IoT is increasingly favored by consumers and business environment. Machine to machine (M2M) has become the fastest growing mobile connection category in IoT application. According to Cisco VNI forecast \cite{b2}, global M2M connections will grow from 780 million in 2016 to 3.3 billion by 2021, going to grow at a compound annual growth rate of 34 percent, and there will be a fourfold growth from 2016 to 2021. The traffic generated by global M2M connections reached 0.14 billion gigabytes per month at the end of 2016, accounting for 2$\%$ of global mobile data traffic. At the end of 2021, M2M traffic will reach 2.45 billion gigabytes per month, making up 5$\%$ of global mobile data traffic.

Traffic management is an important theme to optimize the performance in Internet of Things \cite{b3,b4}. Through the dynamic analysis, forecasting and adjustment of transmission data, the traffic management technology to optimize the performance has been widely used. Existing traffic management technologies rely on closed, rigid existing network architecture designs. The control plane and data plane of this network architecture are tightly coupled, integrated, which hinders current traffic management techniques to provide true differentiated services for large-scale IoT to accommodate fast, growing, uneven, high-speed variable flow mode.

SD-IoT is a new paradigm \cite{b5,b6}, which introduces the software defined framework into the IoT architecture. Under the software defined framework, users can get a dynamic, automated network, which is a bit different than in the past and provides full-virtualization for application requirements. The architecture of the SD-IoT separates the control plane from data plane, which has the following obvious characteristics: visibility, programmability, openness, virtualization, and new flow patterns and characteristics. More importantly, SD-IoT provides a unified global network view that paves the way for inherently flexible, adaptive, customizable traffic control and management technologies of large-scale IoT.

The architecture of SD-IoT consists of two main components: SD-IoT controllers and SD-IoT switches. When an SD-IoT switch receives a new flow, the first packet of the flow is forwarded to a corresponding SD-IoT controller. The SD-IoT controller to compute a forward path for the flow. All SD-IoT switches located on this path install a new forwarding rule. The SD-IoT switch sends the first packet of the new flow to the SD-IoT controller that can cause transmission delays, and the SD-IoT controller calculates the forward path of the flow to generate a processing delay. Even more frightening is that it takes a lot of time to install new forwarding rules on the forward-path switches and is prone to delay peaks. When a large number of new flows are injected into the SD-IoT switch, the control plane and data plane of the SD-IoT architecture all generate significant computing and communication costs. Beacon \cite{b7}, which is the most advanced multi-threaded controller, has a maximum throughput of 12.8 million flow requests per second (Mfrps). The global M2M traffic reached 32407.4 billion bytes per second at the end of 2016 \cite{b2}. Obviously, a single SD-IoT controller has been unable to meet the needs of big data flow for large-scale networks. In order to solve the above-mentioned problems of large-scale IoT, our preliminary work \cite{b8,b9} studied the problem of load balancing of data plane in large-scale software-defined networks. In this paper, we will further study the dynamic load balancing in the control plane of the large-scale SD-IoT. The main contributions of this paper are as follows.

\begin{itemize}
\item A general framework for the large-scale SD-IoT is presented, and it includes two main components: SD-IoT controllers and SD-IoT switches. The SD-IoT controllers use resource pooling technology, which are responsible for the centralized control of network resources. The SD-IoT switches integrate IoT gateway function, and are responsible for data access and forwarding.
\item A vertical control structure for the control plane of SD-IoT is proposed, and it includes the controllers (main controllers) of the main control layer and the controllers (basic controllers) of the basic control layer. The main controllers are responsible for resource management and coordination of the basic control layer, and also provide a northbound interface for the upper application. The basic controllers are responsible for the interaction between the control layer and data forwarding layer.
\item A dynamic load balancing algorithm for the main controllers based on the electoral mechanism is proposed, which selects a main controller called a \emph{Leader} from the main controllers to coordinate the message consistency between the main controllers to ensure that the main controllers can get the same, real-time global network view.
\item A dynamic load balancing model based on balanced delay is deduced, and then a dynamic load balancing algorithm of the basic controller is proposed to reduce the network delay and avoid the traffic load imbalance.
\end{itemize}

The remainder of this paper is organized as follows. Section II summarizes the relevant work and current research trend of the load balancing schemes of the software-defined networking. Section III presents the general framework of the SD-IoT and the vertical control structure of the control plane. Section IV proposes dynamic load balancing algorithms for main controllers and basic controllers. Section V presents the experiment settings and evaluates the performance. Finally, a conclusion is given in Section VI.

\vspace{0.3in}
\section{Related work}

When an SD-IoT switch receives a new flow, it forwards the first packet of the flow to the corresponding SD-IoT controller. The SD-IoT controller calculates and determines the path to forward the flow. The forwarding rules are installed in all SD-IoT switches on the flow path. Obviously, when a large number of new flows are injected into the IoT, the SD-IoT controller will frequently receive and forward the new flow request information and calculate the flow path. At the same time, the installation of the forwarding rule process in the SD-IoT switch will produces a delay peak, easy to cause the Internet traffic load imbalance. Thus, a single centralized controller is easy to become a performance bottleneck. At present, the feasible solutions of the performance bottlenecks can be divided into two categories \cite{b3,b4}: controller load balancing and switch load balancing.

\subsection{Controller Load Balancing}

Controller load balancing refers to the ability to enhance network data processing by using multiple controllers. It can be divided into two levels: horizontal controller load balancing and vertical controller load balancing. Horizontal controllers adopt flat structure. All controllers have the same responsibilities and functions, and the controllers can communicate with each other directly. Vertical controllers adopt hierarchical structure. Controllers in each layer have different responsibility and duty, and upper controllers are responsible for the communication and coordination between lower controllers.

Typical controllers with level structure are HyperFlow \cite{b10}, DIFANE \cite{b11}, Onix \cite{b12},and BalanceFlow \cite{b13}. HyperFlow uses a publishing and subscribing method based on a distributed file system, but it adds additional overhead for subscription management and maintenance. DIFANE uses a core switch instead of a controller, but the core switch increases resource consumption. Both HyperFlow and DIFANE have a logical centralized, physically distributed control plane, and all controllers (or core switches) share a global network view. Onix uses a publishing and subscribing method based on the data structure of the NIB (network information base). Each local controller has an NIB data structure, and the controllers share a copy of the network state with each other, but the controller adds additional subscription management and maintenance overhead. BalanceFlow uses a dedicated controller for load balancing of all controllers and periodic reporting of flow request information. Each controller maintains its own flow request information, but it increases the overhead of the control plane. Onix and BalanceFlow divide a large-scale network into many small networks, and each small network is managed by a local controller.

Typical vertical control structures are Kandoo \cite{b14}, Orion \cite{b15}, SOX and DSOX \cite{b16}, HybridFlow \cite{b17}. Kandoo uses the root controller to control all local controllers, and each local controller manages one or more switches and has a global network view. Orion is similar to Kandoo. SOX, DSOX, and HybridFlow all use logical centralized control plane but physically distributed controller cluster architecture, and each controller cluster shares NIB.

Controller load balancing improves the latency between the switch and the controller in a single controller environment. The level controller architecture provides better recovery capability, but is difficult to manage. The vertical control structure is easier to manage through the upper controller, but there are single points of failure and complete consistency.

However, the load balancing strategy of the control plane has not been exploited to some extent. It needs to solve a series of fundamental problems, which are aimed at finding the optimal number of controllers, deployment location, workload distribution, forwarding path of control messages, and coordinates an optimal balance between the delay performance of control messages and the control costs, and has the statistical characters of network traffic and the diversity of network topology.

Heller \emph{et a}l. \cite{b18} first studied the static deployment of controllers, by calculating the minimum propagation delay to determine the number and location of controllers. Dixit \emph{et al}. \cite{b19} proposed a distributed controller pool architecture that dynamically adjusts the work status of controllers and to balances the real-time workloads of the controllers based on traffic conditions. Bari \emph{et al}. \cite{b20} studied the dynamic supply of controllers in a large-scale local area network, and dynamically changed the controller deployment through the number of real-time flows in the network. Yao \emph{et al}. \cite{b21} studied load balancing based on the capacitated K-center problem, minimizing the maximum delay between the switch and the controller. Jimenez \emph{et al}. \cite{b22} uses the k-critical algorithm to find the minimum number of controllers and the location of the deployment so that the selected controllers are workload balanced. Guo \emph{et al}. \cite{b23} proposed a controller state synchronization strategy based on changes in the load to improve the load balancing performance of multi-controller and multi-domain SDNs. Liao \emph{et al}. \cite{b24} proposed a density-based controller deployment method that divides the network into multiple subnets and configures a controller in each subnet. Huque \emph{et al}. \cite{b25} studied the deployment of dynamic controllers by adjusting the controller deployment location to limit the communication latency and adjusting the number of controllers to support dynamic loads. Hu \emph{et al}. \cite{b13} studied the problem of controller deployment and workload distribution. Schmid and Suomela \cite{b26} proposed a local protocol development algorithm and a localized model of distributed computing. Jim \emph{et al}. \cite{b27} designed the K-Critical algorithm to handle failure and balance load between controllers by finding the minimum number of controllers and its location to establish a controller topology. Ma \emph{et al}. \cite{b28} proposed a load balancing mechanism based on a hierarchical control structure, which meta controllers analyze the resources and utilization in local control planes and optimize the processing performance. However, these efforts focus on finding quantitative or even heuristic results, the lack of deep research to the load balancing mechanism of the control plane in large-scale SD-IoT.

\subsection{Switch Load Balancing}

ECMP \cite{b29} is a load balancing strategy that uses a flow-based hashing method to optimal flow allocation, but two or more long flows are prone to conflict on their hash and share the same output port, resulting in network bottlenecks. Hedera \cite{b30} is an extensible dynamic flow scheduling system that collects statistics information of flows on edge switches. If the flows increases beyond a given threshold rate, it will dynamically compute a suitable path and install the path in the switch. This allows for a balance between the high utilization and minimal scheduling overhead of networks. Mahout \cite{b31} monitors and detects large flows on the terminal host through a mezzanine of the operating system. When the mezzanine detects that the socket's buffer exceeds a given threshold, it will mark the subsequent packet of the flow. The switch forwards these marked packets to the Mahout controller. The Mahout controller calculates the best path for the large flow and installs a specific flow entity in the switch. MicroTE \cite{b32} is similar to Mahout. The above three methods use the central controller to calculate the appropriate flow path, and ECMP routing is used in the switch for a small flow. However, Hedera increases the processing overhead of controllers and switches and the bandwidth overhead of switches. Mahout and MicroTE increase the processing overhead of switches and hosts and the bandwidth overhead of the host.

In order to reduce the number of interactions between controllers and switches, the traffic management of SDNs implements the wildcard rule in OpenFlows witches. Switches can handle the local route of microflows. Controllers only process directional macroflows, especially the quality of service of macroflows is more and more important, such as DevoFlow \cite{b33}. Another approach is using a core switch to complete all the packet processing, such as DIFANE \cite{b11}. However, this approach don't require the controller to participate in the process at all, and reduces the load on the control plane, but increases the burden on the core switch.

However, load balancing policies based on hash ECMP and wildcard rules are static and difficult to adapt to flow dynamics. In \cite{b34}, a decision strategy based on switch migration is proposed, which senses load imbalance through switch migration triggering metrics, establishes a corresponding migration efficiency model, and weighs between migration costs and load balancing rates. Paper \cite{b35} proposed an alternative to the Beacon controller, which collects statistical information of OpenFlow devices in real time. The Beacon controller reroutes the flows according to the queues in the switch to ensure queue balancing in the switch. A load balancing algorithm in SDNs is proposed by collecting the traffic statistics of the switch subset in \cite{b36}. Obviously, the traffic management in SD-IoT needs a dynamic load balancing mechanism, which can dynamically adapt to time-varying network status and fine-grained traffic characteristics, such as traffic burst and arrival time interval.

Recently, paper \cite{b37} proposed a dynamic load balancing method for SDN-based cloud centers, which offers improved flexibility to the task scheduling using of the SDN technology, and completes the real-time monitoring of service node flows and load conditions based on the OpenFlow protocol. The controller can deploy global network resources once imbalance occurs. A constrained optimization particle swarm algorithm based on SDN is proposed, which can effectively reduce the network delay and improve the quality of service of cloud and fog networks in \cite{b38}.

To sum up, in order to avoid the bottleneck of a single centralized controller, traffic management should consider the network traffic load balancing solution. For optimizing the number of SD-IoT controllers, the location of the deployment, the workload distribution, and the control message forwarding path, the traffic equalization solution must facilitate the efficient and accurate acquisition of the traffic statistics of SD-IoT. In order to take full advantage of the  flexible control of SD-IoT and the global view of the characteristics of the SD-IoT, traffic management needs a dynamic load balancing mechanism, which can adapt to time-varying network status, and fine-grained characteristics of flows based on traffic burst and arrival interval can be adjusted. The above studies have shown that the effective dynamic load balancing strategy of vertical controllers in SD-IoT is especially important for large-scale IoT.

\section{Vertical Control Structure for SD-IoT}

\subsection{SD-IoT Framework}

\begin{figure}[!t]
\centering
\includegraphics[width=6in]{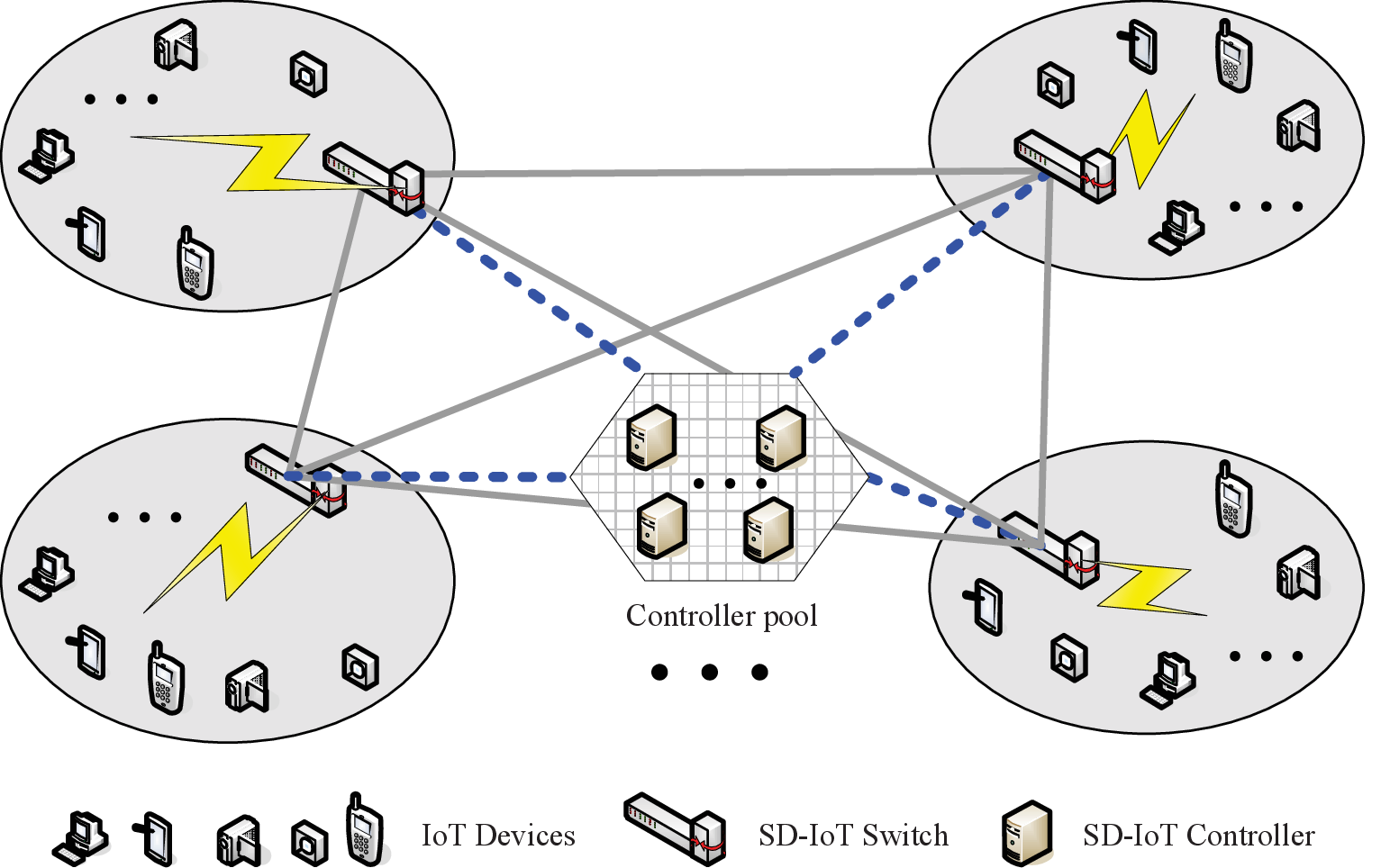}
\caption{A generic framework for SD-IoT.}
\label{fig1}
\end{figure}

In this subsection, we describe a generic framework for SD-IoT, as shown in Fig. \ref{fig1}. The generic framework for SD-IoT can be divided into three layers: application layer, control layer and infrastructure layer. The application layer consists of IoT servers, which connect to SD-IoT controllers through the Internet and provide a variety of applications and services through APIs. The control layer is composed of SD-IoT controllers, which run the distributed operating system, and provides a logical centralized control and view for the network data forwarding in the distributed IoT. SD-IoT controllers use the resource pooling technology, and the use of controller resources in the controller pool can be dynamically adjusted based on the real-time conditions of IoT switch resources. The infrastructure layer is composed of SD-IoT switches, which integrate the functions of IoT gateways and SDN switches, and access different actuating and sensing devices of IoT, such as cameras, digital cameras, smart phones and personal computers by controlling the interface of data plane in SD-IoT.

\subsection{Vertical Control Structure}

\begin{figure}[!t]
\centering
\includegraphics[width=5in]{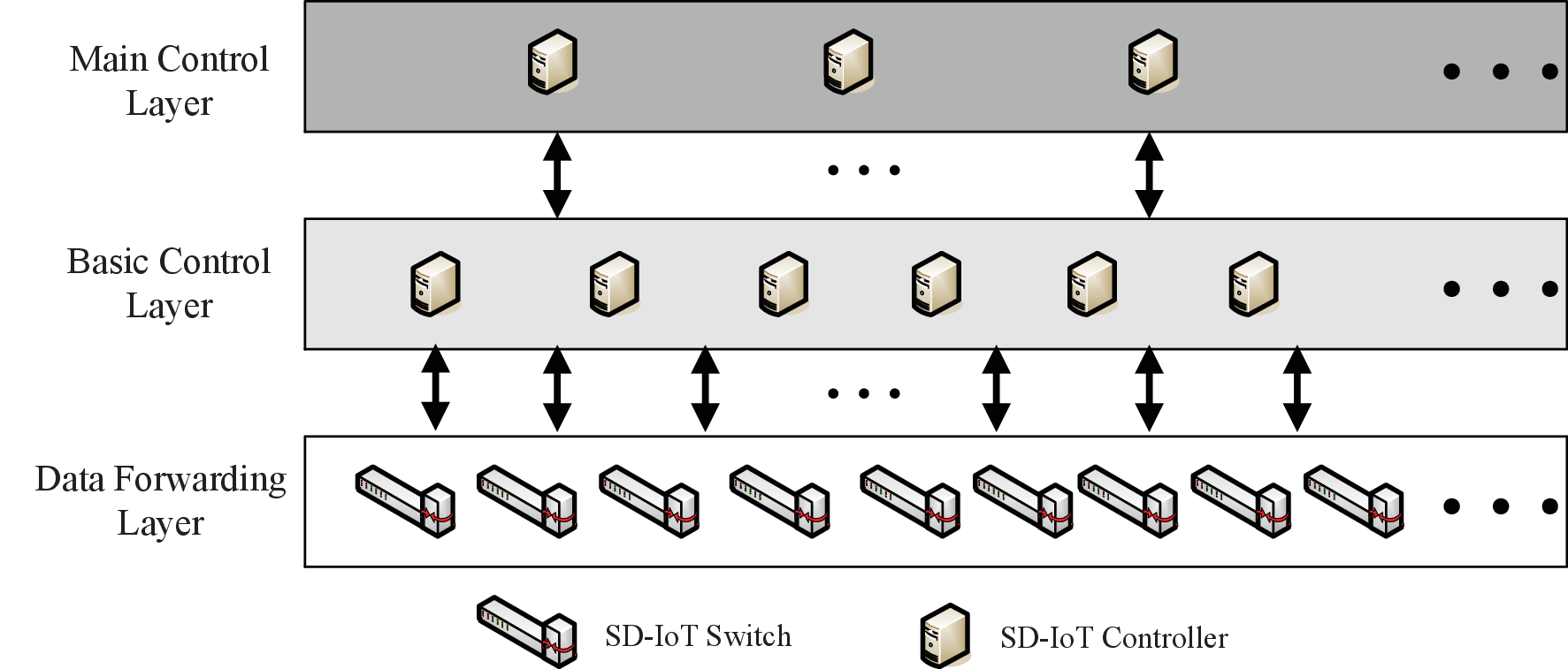}
\caption{Vertical control structure of the controller pool.}
\label{fig2}
\end{figure}

In the above proposed framework of SD-IoT, the controller pool is designed as a vertical control structure, as shown in Fig. \ref{fig2}. The vertical control structure includes the main control layer and basic control layer. The main control layer interacts with the application layer upwards, and interacts with the basic control layer downwards. The basic control layer interacts with the data forwarding layer downwards.

The controllers of the main control layer are called the main controllers. Each of the main controllers manages the base controllers in the base control layer, while the other basic controllers are used as resources in reserve. The main controllers are responsible for resource management and coordination of the base control layer, and also provide a northbound interface for the upper application. In the main controllers, a \emph{Leader} is generated by the election mechanism. The \emph{Leader} can obtain the global network topology information, control the main controller, and coordinate the basic controllers.

The basic controllers are responsible for device resource management in a region of IoT, as shown in Fig.\ref{fig1}. The IoT devices in the same region can communicate through the basic controller of the region. Each switch will be one of the base controllers as a \emph{Master} controller, and others as \emph{Slave} controllers. The IoT devices in different areas communicate with each other through a main controller. The basic controller communicates with the switch of the data forwarding layer, regularly sends Packet\_out messages to the switch, and obtains the information of the switch through feedback Packet\_in messages. The basic controller submits its own control information by interacting with the main controller so that the main controller can obtain the entire network global view. Network load balance is achieved through the coordination of messages between main controllers and the dynamic load balancing strategy in the base controller layer.

Compared with the horizontal controller structure, the vertical control structure is easy to manage, but also solves a series of problems caused by single point failure of the existing vertical control structure. The problems include the incongruity of underlying controllers, non-global topology, load imbalance, the inconsistencies of main controllers.

\section{Dynamic load balancing algorithm for vertical control}

In this section, a dynamic load balancing strategy of vertical controllers is proposed using the proposed SD-IoT framework. It consists of two parts: a dynamic load balancing algorithm of the main control layer and a dynamic load balancing algorithm of the basic control layer. The former is responsible for the single point of failure of the main controller and the consistency of all controller messages. The latter is responsible for the load balancing of the basic controllers.

\subsection{Dynamic load balancing algorithm for the main control layer}

Existing vertical control structures are generally used with a controller or super controller \cite{b14,b17}. However, they are easy to prone to a single point of failure. Multiple super controllers can solve the problem of single point of failure \cite{b16}, but there is the inconsistency of messages between the main controllers. In order to solve this problem, it is necessary to construct a set of effective message consistency mechanism in the multi-controller structure. PAXOS \cite{b39} is a consistency algorithm based on the information transfer model, which is considered to be the most effective of similar algorithms. In this section, we proposed a dynamic load balancing algorithm for the main control layer based on PAXOS algorithm.

\subsubsection{Algorithm ideas}

A dynamic load balancing algorithm for the main control layer proposed in this section is to use a set of ideas to select a main controller as a \emph{Leader} from multiple main controllers to ensure that multiple main controllers interact with each other, and the \emph{Leader} manages execution and replacement information is passed to other main controllers so that all network views held by all the main controllers are quickly consistent, and the single point of failure is also optimized. Similar to the PAXOS algorithm, the proposed algorithm consists of three roles: a \emph{Proposer}, an \emph{Acceptor} and a \emph{Learner}. The \emph{Proposer} is responsible for proposing proposals. The \emph{Acceptor} votes on the nomination proposal. The \emph{Learner} collects the proposal accepted by each \emph{Acceptor}. All the main controllers can be any of three roles.

\subsubsection{Algorithm design}
Assume \emph{n} controllers in a main control layer are sent to their respective \emph{Proposer} a message and claim to be a \emph{Leader}, the \emph{Proposer} sends a proposal with their own $<$\emph{K}, \emph{V}$>$ message to all \emph{Acceptors}. According to \emph{K}, all \emph{Acceptors} will complete the update of \emph{K} and promise to ensure that the proposal is rejected less than the updated \emph{K} value, at the same time pass the proposal $<$\emph{K}, \emph{V}$>$. When no more than half of the feedback was received by the \emph{Proposer}, it will update itself of \emph{K}, and continue to give proposal messages from all \emph{Acceptors}. Finally, $<$$K_N$, $Controller_N$$>$ is the proposal to pass the resolution, and $K_N$ is the largest number and $Controller_N$ is the largest numbered controller. At this point, the \emph{Learner} perceives the adoption of the $<$$K_N$, $Controller_N$$>$ proposal and learns the proposal. At the end of the process, $Controller_N$ is elected as a \emph{Leader}. Once the Leader fails, it will return to zero to launch a new election.

\subsubsection{Algorithm implementation}

The dynamic load balancing algorithm proposed in this section is deployed in the main controllers, so that when multiple main controllers interact with each other, the Leader manages the execution and replacement information to pass to other main controllers. The main implementation process of the dynamic load balancing algorithm in the main control layer is as follows in Algorithm \ref{alg1}.

\begin{algorithm}[H]
\caption{\textbf{:}~~A dynamic load balancing algorithm for the main control layer.}
\label{alg1}
\begin{algorithmic}[1]

\REQUIRE~The number $N$ of controllers.
\ENSURE~The \emph{Leader} and election time.
\FOR {$all$ $controllers$}
\STATE \emph{All controllers send a message to the Proposer to elect the leader and generate the value of K based on the time stamp.}

\FOR {$<$$K$,$V$$>$ $of$ $all$ $Proposers$ $\rightarrow$ $Pcceptors$}
\IF {$new(K)>old(K)$}
\STATE $Update(new(K),controller_N)$
\ENDIF

\IF {\emph{any of Proposers receives the number Acceptors returned more than half}}
\STATE \emph{this Proposer will update its own K}
\STATE \emph{the new K is used to send election massages to other Acceptor}
\ENDIF

\ENDFOR
\STATE \emph{Gain the Leader}
\ENDFOR
\STATE \emph{Gain the election time}
\end{algorithmic}
\end{algorithm}

The dynamic load balancing algorithm proposed in this section mainly uses the idea of the set, and solves the problem of single point failure of the existing vertical control structure, and ensure that the global views of the network in the main controllers are consistent with each other.

\subsection{Dynamic load balancing algorithm for basic control layer}

In the vertical control structure of the SD-IoT, the request messages sent by the switch on the basic control layer to controllers are too large, and it can make huge burden on the corresponding controller and result in network delay. In a worse case scenario, it will lead to the base controller failures and network collapse. Other base controllers may be in an idle state. In order to optimize the above problems in the basic control layer, we propose a dynamic load balancing algorithm based on balanced delay for the basic controllers in this section. When a base controller fails or the single base controller is overloaded and the other base controllers are in the idle state, the algorithm is used to make the switches managed by the base controller of the faulty or the overload manage the other switches migrate to other base controllers, and achieves the load balancing of the basic control layer.

\subsubsection{Load balancing model}

In a vertical control architecture, each switch is connected to one or more controllers, and one controller controls multiple switches. Assume that the number of SD-IoT base controllers in the control plane is \emph{n}, we have $C=\{C_i\}_{i=1,2,...,n}$, where $C$ indicates a set of controllers and $C_i$ is the \emph{i}th controller. Assume that the number of the SD-IoT switches in the data plane is \emph{m}, we have $S=\{S_j\}_{j=1,2,...,m}$, where $S$ indicates a set of switches and $S_j$ is the \emph{j}th switch. Obviously, the control relationship between \emph{n} base controllers and \emph{m} switches can be represented by a matrix $Q_{nm}$ of \emph{n} rows and \emph{m} columns, as shown in Equ.(\ref{equ1}).
\begin{equation}
\label{equ1}
Q_{nm}=
\left[
  \begin{array}{cccc}
    q_{11} & q_{12} & ... & q_{1m}\\
    q_{21} & q_{22} & ... & q_{2m}\\
    ... & ... & ... & ...\\
    q_{n1} & q_{n2} & ... & q_{nm}\\
  \end{array}
\right]
\end{equation}
where $q_{ij}=0$ indicates that the \emph{j}th switch $S_j$ is not controlled by the \emph{i}th base controller $C_i$; $q_{ij}=1$ indicates that the \emph{j}th switch $S_j$ is controlled by the \emph{i}th base controller $C_i$.

Assume that the number of the Packet\_in packets sent by the \emph{j}th switch $S_j$ to the \emph{i}th base controller $C_i$ at time \emph{t} is $p_{ij}$, we have
\begin{equation}\label{equ2}
  p_{ij}=q_{ij}f_{ij}(t)
\end{equation}
where $f_{ij}(t)$ is the rate at which the \emph{j}th switch $S_j$ sends the packet.

Obviously, during the interval [0, $T$], the total amount of Packet\_in packets sent by the \emph{j}th switch $S_j$ to the \emph{i}th base controller $C_i$ is $P_{ij}$, and we have
\begin{equation}\label{equ3}
\begin{aligned}
  P_{ij}&=\int_0^T p_{ij}dt\\
  &=\int_0^T q_{ij}f_{ij}dt
\end{aligned}
\end{equation}

So, the total number of Packet\_in packets processed by the \emph{i}th base controller $C_{ij}$ from \emph{m} switches: $S_1, S_2, ..., S_m$ at time $t$ is expressed by
\begin{equation}\label{equ4}
p_i=\sum_{j=1}^{m}(q_{ij}f_{ij}(t))
\end{equation}

During the interval [0, $T$], the total number of Packet\_in packets processed by the \emph{i}th base controller $C_{ij}$ from \emph{m} switches: $S_1, S_2, ..., S_m$ is expressed by
\begin{equation}\label{equ5}
P_i=\int_0^T \sum_{j=1}^{m}(q_{ij}f_{ij}(t))dt
\end{equation}

Thus, the total number of Packet\_in packets processed by the \emph{n} base controllers in the basic control layer from the \emph{m} switches in the data plane layer can be expressed by
\begin{equation}
\label{equ6}
\begin{aligned}
P&=\sum_{i=1}^{n} P{i}\\
&=\sum_{i=1}^{n}(\int_0^T \sum_{j=1}^{m}(q_{ij}f_{ij}(t))dt)\\
&=\int_0^T \sum_{i=1}^{n} (\sum_{j=1}^{m}(q_{ij}f_{ij}(t)))dt
\end{aligned}
\end{equation}

In vertical control structure of the large-scale SD-IoT, once the edge switch receives a new flow, the edge switch will forward the first packet of the flow to the corresponding base controller. The base controller calculates and determines the path to forward the flow. All switches on the flow path will install forwarding rules. Obviously, when a large number of new flows into the IoT, the basic controller will frequently receive and forward the new flow request information and calculate the flow path. The switches on the path will frequently install forwarding rules. It will produce a peak of the network delay, and cause the load imbalance of the IoT.

Network delay is mainly composed of processing delay, queuing delay, transmission delay and sending delay. In general, the processing delay and sending delay of the controller and switch can be considered constant. Therefore, the network delay depends mainly on the queuing delay and transmission delay. From the queuing model M/M/1, it can be concluded that the queuing delay $T_{j,w}$ of a Packet\_in packet sent by the \emph{j}th switch $S_j$  in the \emph{i}th base controller $C_i$ can be expressed as
\begin{equation}\label{equ7}
  T_{i,w}=\frac{\lambda_i}{\mu_i(\mu_i-\lambda_i)}
\end{equation}
where $\lambda_i=\frac{P_i}{T}$ is the arrival rate of packets, that is, the average value of Packet\_in packets arriving at the \emph{i}th base controller $C_i$ in unit time; $ \mu_i=\frac{P_i}{T_d}$ is the service rate of controllers, that is, the average rate of the \emph{i}th base controller $C_i$ precessing Packet\_in packets.

Assuming that the transmission delay of packets transmitted by the \emph{j}th switch $S_j$ of the data forwarding layer to the \emph{i}th base controller $C_i$ of the control layer is $T_{i,c}$, which can be obtained by calculating the maximum delay of all effective shortest paths between the switch $S_j$ and the base controller $C_i$ \cite{b42}, we have
\begin{equation}\label{equ8}
  T_{i,c}=\max_{S_j\in S}\min_{C_i\in C}d(S_j,C_i)
\end{equation}

As a result, the total time $L_i$ of the \emph{j}th switch receiving a new flow, the \emph{i}th controller calculating the path and the switches on the path installing forwarding rules is expressed by
\begin{equation}\label{equ9}
  L_i=T_s+T_d+T_{i,w}+T_{i,c}
\end{equation}
where $T_s$ is the processing delay and $T_d$ is the transmission delay.

Thus, the sum of the delay between the base control layer and the data forwarding layer is
\begin{equation}\label{equ10}
\begin{aligned}
  L&=\Sigma_{i=1}^{n}L_i\\
  &=\Sigma_{i=1}^{n}(T_s+T_d+T_{i,w}+T_{i,c})\\
  &=\Sigma_{i=1}^{n}(T_s+T_d+\frac{\lambda_i}{\mu_i(\mu_i-\lambda_i)}+\max_{S_j\in S}\min_{C_i\in C}d(S_j,C_i))
\end{aligned}
\end{equation}

In the SD-IoT vertical control structure, the workload from basic controllers can be approximated as the amount of Packet\_in packet requests \cite{b41}. Therefore, the load balancing model of basic controllers can be expressed as
\begin{equation}\label{equ11}
  B=\{Q_{nm},L,P\}
\end{equation}
where $Q_{nm}$, $L$ and $P$ are given by Equ.(\ref{equ1}), (\ref{equ6}) and (\ref{equ10}), respectively.

In fact, if you can reduce the network delay, it is easy to achieve load balancing. That is, the load balancing model can be converted to the lowest delay model, this is
\begin{equation}\label{equ12}
  B=\{Q_{nm},L\}
\end{equation}

When the switch's requests for the base controller are too large or the base controller fails, the switch is mapped to a different base controller for balancing delay \cite{b41}. It can be seen from Equ.(\ref{equ10}) and Equ.(\ref{equ12}) that the network congestion caused by the excessive load of basic controllers can be alleviated by reducing the queuing delay and transmission delay.

\subsubsection{Dynamic load balancing algorithm for basic controllers}

In this section, we present a dynamic load balancing algorithm for basic controllers based on balanced delay. The proposed algorithm obtains the network topology $G(S, C)$ of switches in the data forwarding layer through basic controllers. Assuming that the threshold for the Packet\_in packet processed by a base controller in unit time is $P_{th}$, which can also be referred to as the load peak of base controllers. When the total number of the Packet\_in packets which the \emph{i}th base controller $C_i$ processes from the \emph{m} switches is greater than or equal to the load peak $P_{th}$ of the \emph{i}th base controller $C_i$, and the request amount of Packet\_in packets which the $j$th switch $S_j$ sends to the \emph{i}th base controller $C_i$ is $p_i$, and $p_i>\frac{P_{th}}{m}$. At the same time, the total value of the queuing delay and transmission delay exceeding the initial value is greater than the total value of the transmission delay and processing delay used to retransmit the Packet\_in packet to the unoccupied basic controllers at the shortest distance, and the corresponding basic controller changes Slave into Master, that is, $q=1$. While the portion of the Packet\_in packets beyond the original base controller is handed over to the idle base controller which has an minimum distance. Otherwise, the packets continue to wait for the base controller to handle until the iteration is complete. When the base controller fails, that is, $q=0$, the switch through the $G(S, C)$ will make the idle Slave with a shortest distance as the Master, that is, $q=1$, and it repeats the above steps. The specific implementation code of the proposed algorithm is as shown in Algorithm \ref{alg2}.

\begin{algorithm}[!t]
\caption{\textbf{:}~~Dynamic load balancing algorithm for the base control layer.}
\label{alg2}
\begin{algorithmic}[1]

\REQUIRE~The initial value of $Q_{nm}$, the rate $p_{ij}$ of Packet\_in packets.
\ENSURE~The value of $Q_{nm}$.
\FOR {$S_j\in C_i$}

\IF {$C_i$ $goes$ $down$}
\STATE $q_{ij}=0$
\STATE $S_j\rightarrow C_{\min d(s,c)}$
\STATE $q_{\min d(s,c)j}=1$
\ENDIF

\IF {$p_{ij}>\frac{P_{th}}{m}$ and $T_s+T_w(p_{ij}-\frac{P_{th}}{m})>T_c(p_{ij}-\frac{P_{th}}{m})+T_d$}
\STATE \emph{Add and control the controller with the smallest distance in the topology for the switch}.
\STATE $q_{\min d(s,c)j}=1$
\STATE \emph{The packet-in message exceeding the threshold set by the controller is processed to the new controller}.
\ENDIF
\ENDFOR
\STATE \emph{Get the value of the newly controlled condition}.
\end{algorithmic}
\end{algorithm}

The dynamic load balancing algorithm proposed in this section transforms the load balancing problem into a network latency problem, which avoids load imbalance by reducing the queuing delay and transmission delay.

\section{Performance Evaluation}

\subsection{Experimental environment}

\begin{figure}[H]
\centering
\includegraphics[width=3.5in]{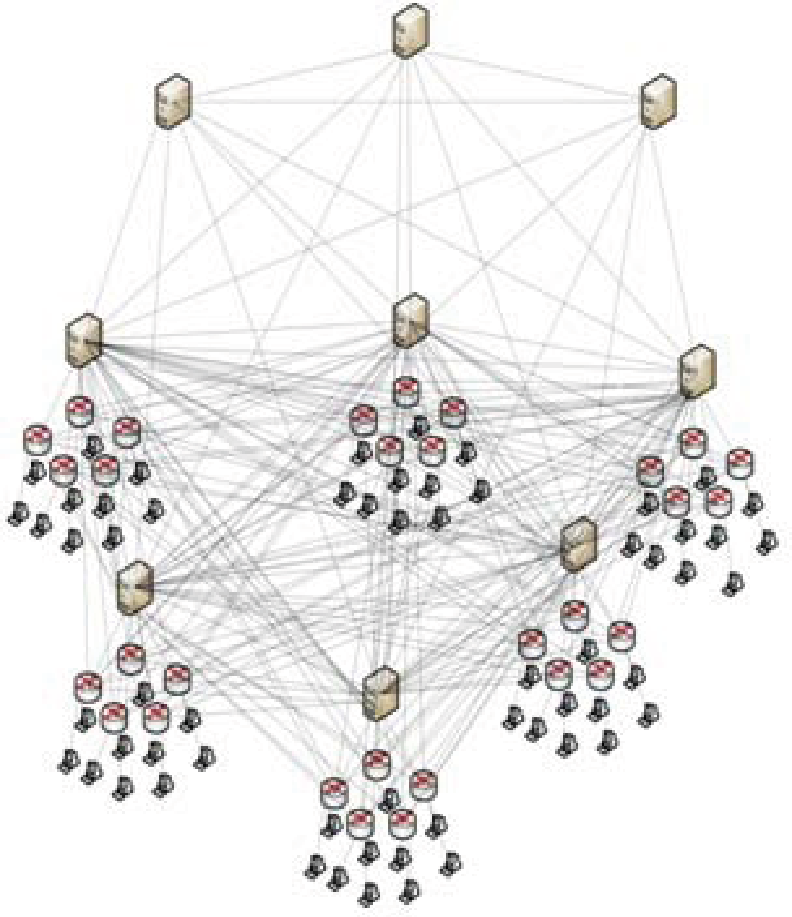}
\caption{A scenario of the SD-IoT.}
\label{fig3}
\end{figure}

In order to evaluate the effectiveness of the proposed SD-IoT framework and the performance of the proposed dynamic load balancing algorithms for vertical control structure of the SD-IoT, we set up the following experimental platform. The hardware environment of the platform includes: Intel core i7-5960X, DDR4 213316G, SSD 250G + 1T HDD 7200 and Intel PRO/1000 MT. The software environment includes: Ubuntu14.04LTS, OpenDaylight, Mininet2.2, VMware Workstation 12 pro and RabbitMQ 3.6.3. The main controllers and base controllers of the control layer in SD-IoT all use OpenDaylight as controllers based on the OSGi architecture. The switches of the data forwarding layer adopt a simulation platform: Mininet. The communication between the main controllers and base controllers is implemented using JGroups in OpenDaylight. The communication between the base controllers and switches is implemented using the OpenFlow protocol and OF-CONFIG protocol. Fig.\ref{fig3} shows a scenario of the SD-IoT. There are nine controllers, including three main controllers and six basic controllers. Each of main controllers connect to six basic controllers, and each of base controllers manages five switches in the local region and connects with other regional switches. We deploy Algorithm \ref{alg1} presented in this paper in the main controllers, and deploy Algorithm \ref{alg2} presented in this paper in the base controllers.

\subsection{Performance analysis}

\subsubsection{Performance evaluation indicators}

For the dynamic load balancing algorithm of main controllers, we use the time electing a \emph{Leader} as an evaluation index. In order to evaluate the performance of the dynamic load balancing algorithm of base controllers, we use a terminal in Mininet to send packets to another address in the basic control layer. The the packet passes a switch in Mininet. If no match is found in the flow table, the switch sends a Packet\_in packet to the controller in such a way that the switch generates a Packet\_in packet in unit time to provide the load pressure to the controller of the base control layer.

Assuming that the average delay of the switch sending requests to the base controller responding to requests is $\overline{T}$, the standard deviation is
\begin{equation}\label{equ13}
  \sigma=\sqrt{\frac{\sum_{i=1}^n(T_i-\overline{T})^{2}}{n}}
\end{equation}

Therefore, in the basic control layer, the response time of the controllers, the standard deviation of the response time of the controllers and the CPU utilization rate of the controllers will be used as the evaluation index of the dynamic load balancing algorithm of basic controllers.

\subsubsection{Results analysis}

In order to evaluate the performance of the dynamic load balancing algorithm proposed in this paper. We deploy the dynamic load balancing algorithm of the main controllers in a variety of different scenarios. The results are obtained by duplicating the experiment multiple times.

\begin{figure}[H]
\setcaptionwidth{2.65in}
\begin{minipage}[t]{0.5\linewidth}
\centering
\includegraphics[width=3.35in]{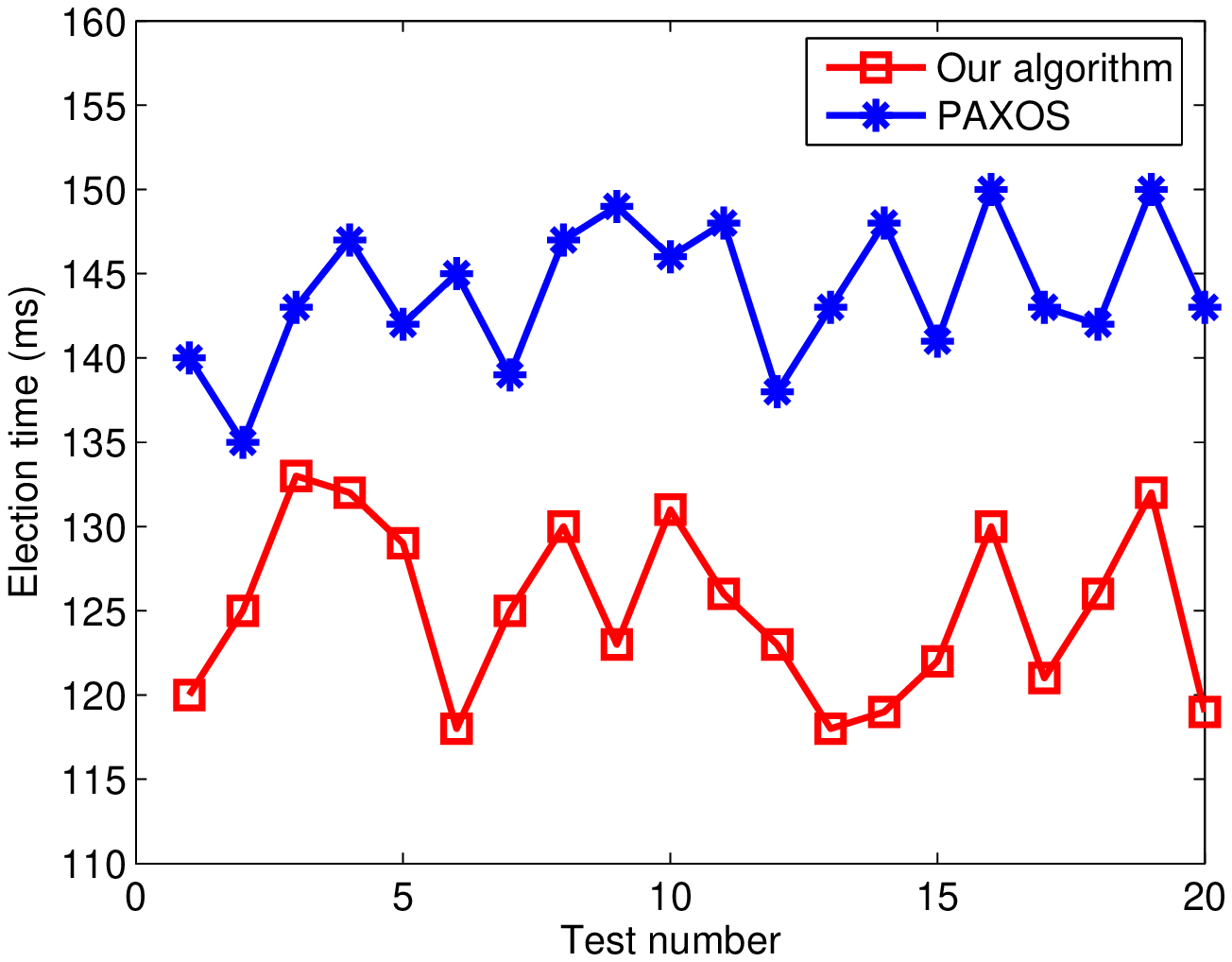}
\caption{Election time with three controllers.}
\label{fig4}
\end{minipage}
\begin{minipage}[t]{0.5\linewidth}
\centering
\includegraphics[width=3.35in]{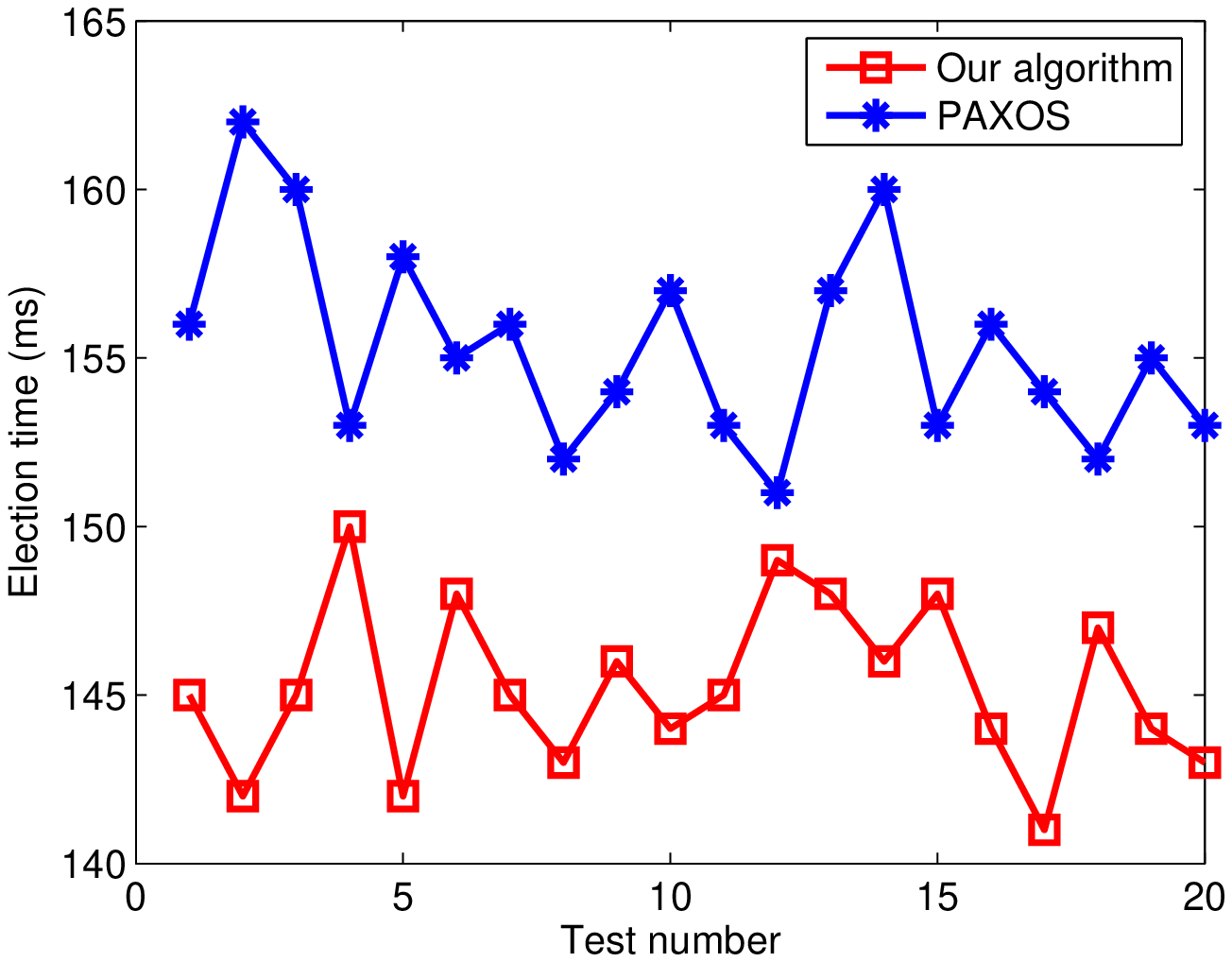}
\caption{Election time with five controllers.}
\label{fig5}
\end{minipage}
\end{figure}

The impact of test number on the time spent on electing a \emph{Leader} from three main controllers is given in Fig.\ref{fig4}. As can be seen from Fig.\ref{fig4}, the average value of the election time of our algorithm is significantly less than that of the PAXOS algorithm. The average time spent on electing a \emph{Leader} from three main controllers based on our algorithm is is about 125 ms, and the time based on the PAXOS algorithm is about 144 ms, and it is reduced by 13 percent. Our algorithm adopts the idea of set, which reduces the time of two stages in PAXOS. The impact of test number on the time spent on electing a \emph{Leader} from five controllers is given in Fig.\ref{fig5}. We have the same results. The average time spent on electing a \emph{Leader} based on our algorithm is about 145 ms, and the average time using the PAXOS algorithm is about 155ms.

In order to evaluate the performance of the dynamic load balancing algorithm for basic controllers proposed in this paper. We deploy Algorithm \ref{alg1} in three main controllers, and deploy Algorithm \ref{alg2} in six base controllers.

\begin{figure}[H]
\setcaptionwidth{2.65in}
\begin{minipage}[t]{0.5\linewidth}
\centering
\includegraphics[width=3.35in]{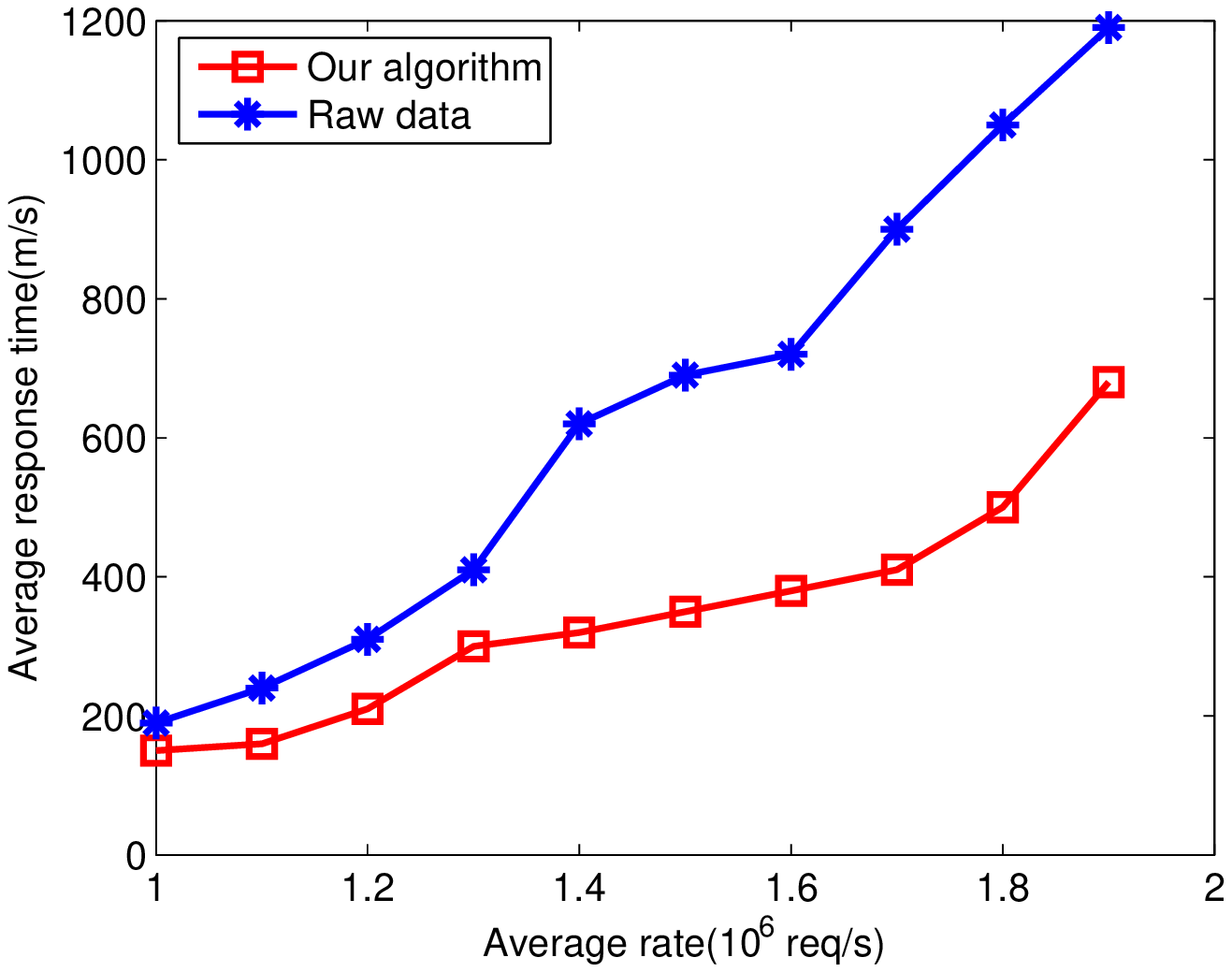}
\caption{Average response time vs. average rate.}
\label{fig6}
\end{minipage}
\begin{minipage}[t]{0.5\linewidth}
\centering
\includegraphics[width=3.35in]{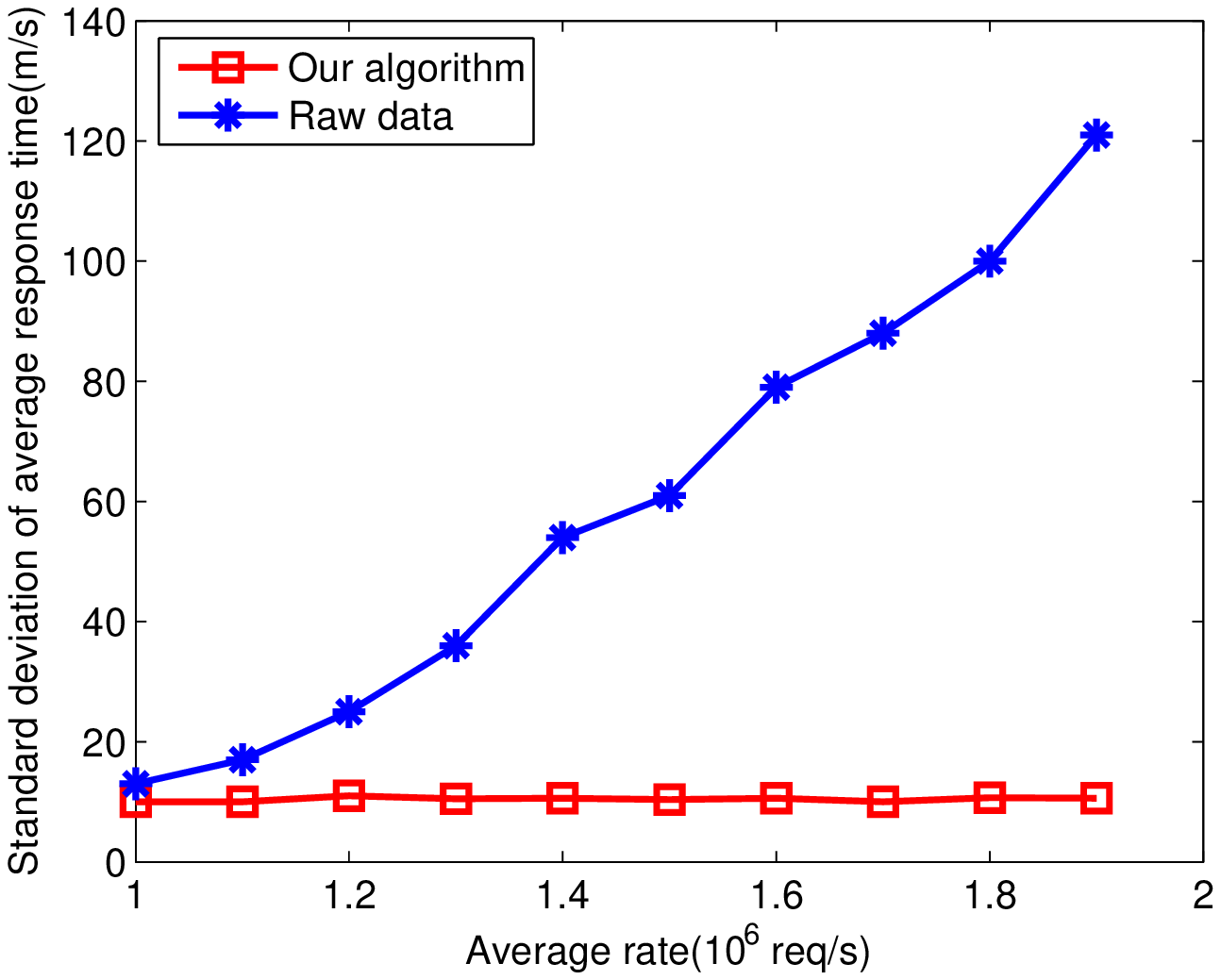}
\caption{Standard deviation of average response time vs. average rate.}
\label{fig7}
\end{minipage}
\end{figure}

The impact of average request response time of base controllers on the rate at which a switch sends Packet\_in packets. As can be seen from Fig.\ref{fig6}, the greater the rate at which the switch sends Packet\_in packets, the more data packets received by the base controllers, and the average response time of the base controllers is gradually increased. The performance of the dynamic load balancing algorithm for basic controllers proposed in this paper is far superior to that of the original data without deploying load balancing algorithm. With the increase of Packet\_in packet rate, the advantages of the proposed algorithm are becoming more and more obvious. The impact of standard deviation of average request response time of base controllers on the rate at which a switch sends Packet\_in packets. It can be seen from Fig.\ref{fig7} that the standard deviation of the controller response time increases rapidly as the Packet\_in packet rate increases without the deployment of the load balancing algorithm. In the case of deploying the load balancing algorithm for basic controllers, the standard deviation of the controller response time is almost unaffected by the Packet\_in packet rate of the switches. The dynamic load balancing algorithm for basic controllers balances the request response time for the switches to send Packet\_in packets to base controllers.

It can be seen that the standard deviation of the controller's response time depends only on the delay of switches and controllers, regardless of the Packet\_in packet rate of switches. In the absence of the deployment of the load balancing algorithm, due to the fact that the load of each controller is different, the response time is different and the average response time of the entire controller is relatively large. Obviously, the proposed algorithm is very suitable for load balancing of large-scale IoT.

\begin{figure}[H]
\setcaptionwidth{2.65in}
\begin{minipage}[t]{0.5\linewidth}
\centering
\includegraphics[width=3.35in]{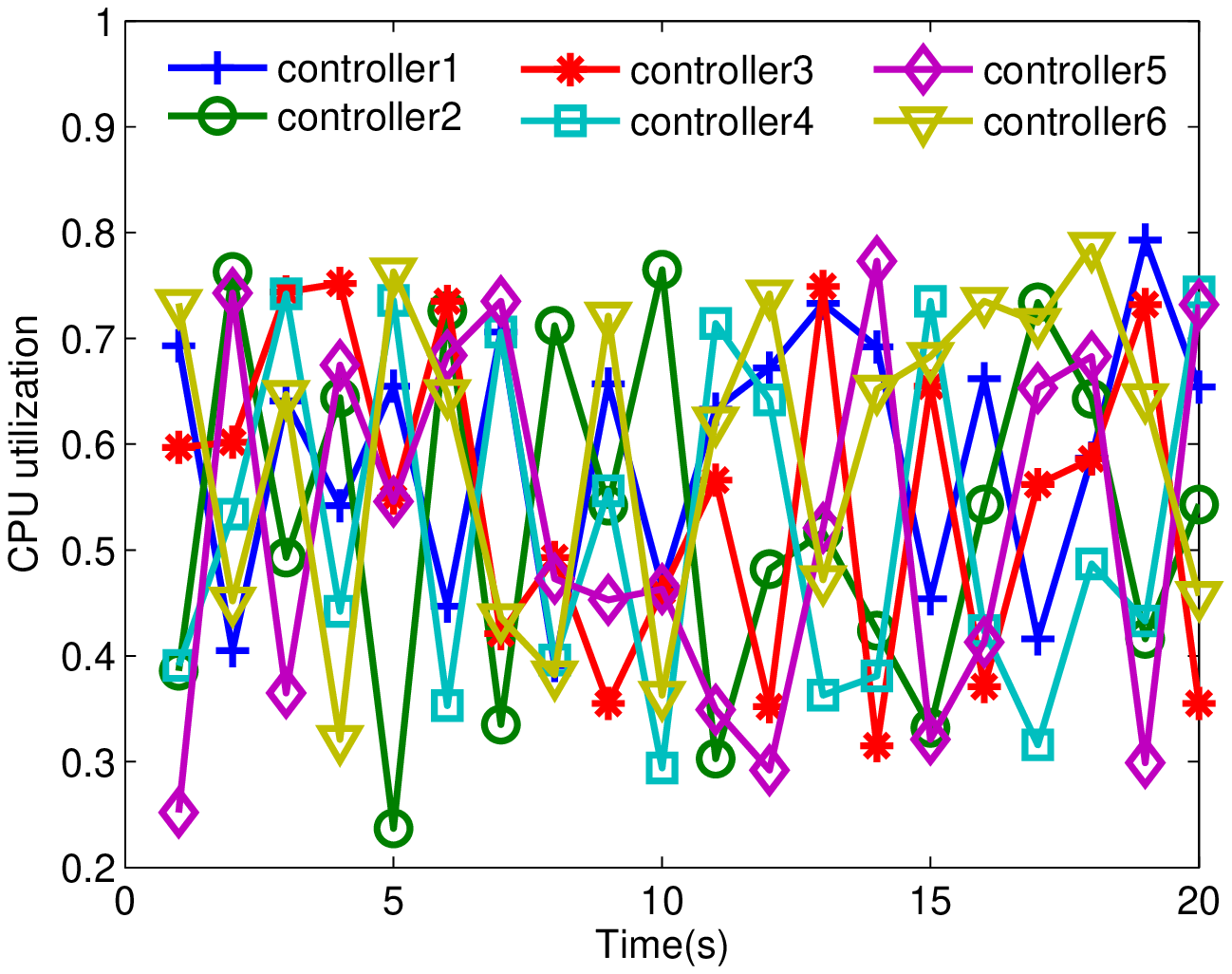}
\caption{CPU utilization under the load balancing algorithm (case 1).}
\label{fig8}
\end{minipage}
\begin{minipage}[t]{0.5\linewidth}
\centering
\includegraphics[width=3.35in]{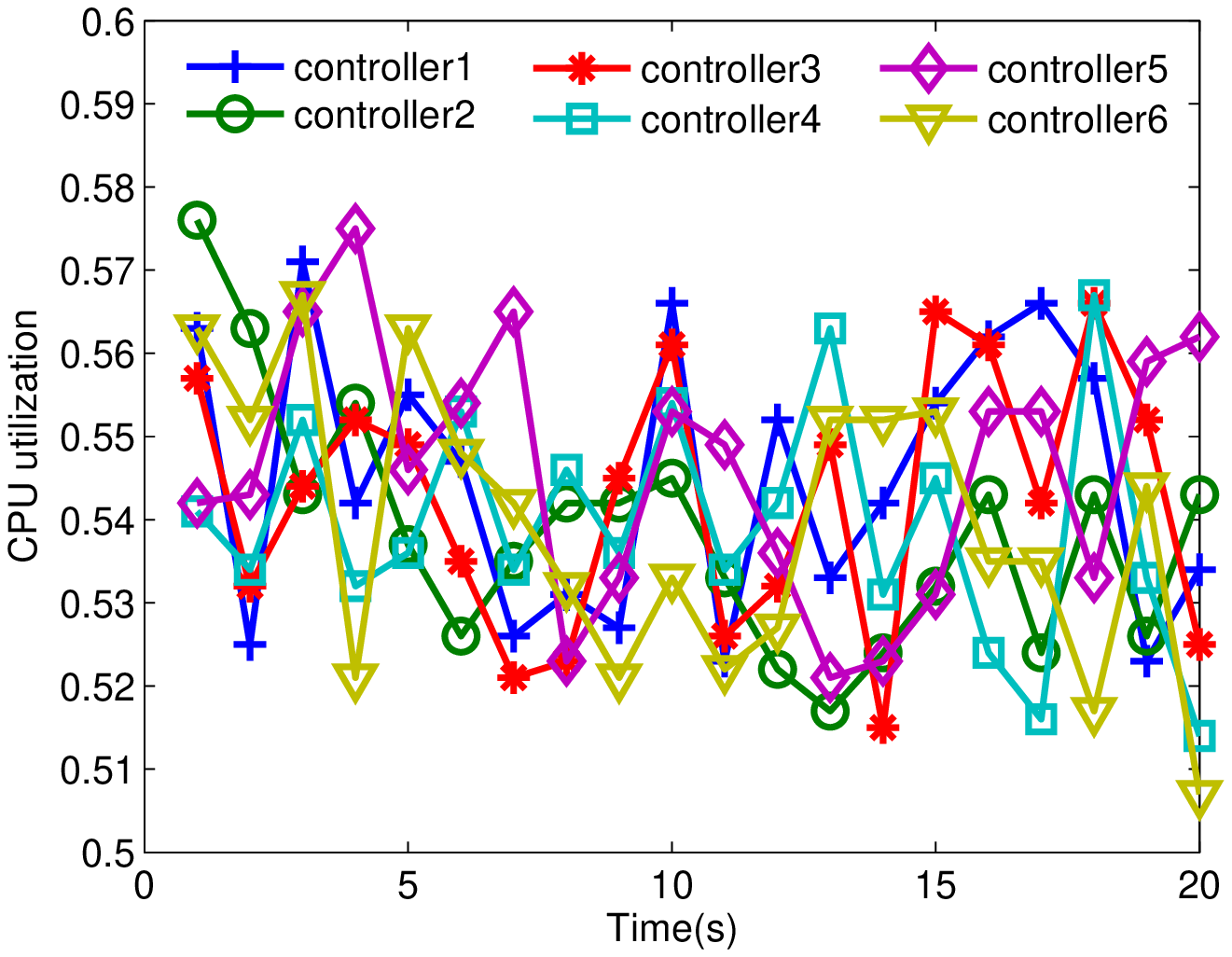}
\caption{CPU utilization without dynamic load balancing algorithm (case 2).}
\label{fig9}
\end{minipage}
\end{figure}

The impact of CPU utilization for configuring six base controllers without deploying the dynamic load balancing algorithm presented in this paper (case 1) is given in Fig.\ref{fig8}. As can be seen from Fig.\ref{fig8}, in the case of the CPU utilization of basic controllers changes dramatically over time fluctuations. The impact of CPU utilization for configuring six base controllers with deploying the dynamic load balancing algorithm presented in this paper (case 2) is given in Fig.\ref{fig9}. As can be seen from Fig.\ref{fig9}, the CPU utilization of basic controllers fluctuates with slighter rate.

\begin{figure}[!t]
\centering
\includegraphics[width=5.5in]{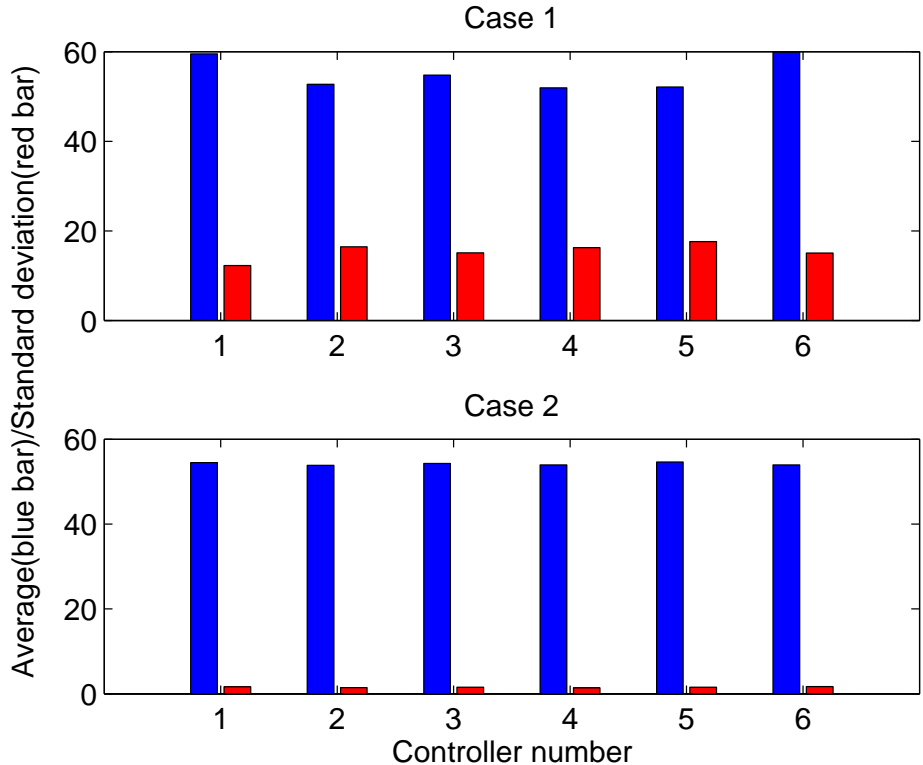}
\caption{Average and standard deviation of the CPU utilization of basic controllers.}
\label{fig10}
\end{figure}

The average (blue bar) and standard deviation (red bar) of the CPU utilization of basic controllers in both cases are given in Fig.\ref{fig10}. In both cases, the average of the CPU utilization of the six base controllers is close. In case 1, the average CPU utilization of the six base controllers is 55.15$\%$. In case 2, the average CPU utilization of the six base controllers is 54.18$\%$. However, the standard deviation of the CPU utilization of the base controllers in case 2 is much less than that in case 2. The average standard deviation of the CPU utilization of the six basic controllers is about 15.44$\%$ in case 1, but it is about 1.55$\%$ in case 2. Obviously, the CPU utilization in case 1 has better stability than that in case 2, that is, the load is more balanced. The load balancing algorithm of basic controllers is based on the balance delay to select a controller, and achieves the purpose of load balancing.


To sum up, we can see that the dynamic load balancing algorithm for main controllers proposed in this paper can make main controllers to quickly synchronize the information of the global network view. The load balancing algorithm for basic controllers proposed in this paper can ensure that basic controller resources are distributed evenly to achieve the effect of load balancing.

\section{Conclusions}
This paper describes a general framework for SD-IoT. The control plane of the framework uses a vertical control architecture, which is divided into the main control layer and the basic control layer. The main controllers in the main control layer coordinate and manage the basic control layer, and the basic controllers in the basic controller layer manage and control the data forwarding layer of switches. In the main control layer, we design an algorithm for electing a controller as a \emph{Leader}, which is used to coordinate and manage the main controllers, in order to achieve the dynamic load balancing of main controllers. In the basic control layer, we design a dynamic load balancing algorithm based on balanced delay, which is used to process Packet\_in packets of witches in the data forwarding layer to ensure the dynamic load balancing of the base control layer. The experimental results show that the proposed SD-IoT framework and dynamic load balancing algorithms in the vertical plane are effective. The work would assist and support traffic control and management of large-scale IoT.

\vspace{0.3in}
\section{Acknowledgements}
This work was supported in part by Natural Science Foundation of China (Grant No. 61572191 and 61402170) and the Research Foundation of Education Bureau of Hunan Province of China (Grant No.17A130).

\end{document}